\documentclass[aps,twocolumn,prd,showpacs,nofootinbib,showkeys]{revtex4}

\usepackage{graphicx}
\usepackage{amsmath}
\usepackage{amssymb}
\newcommand{\dd}{\mathrm{d}} 
\newcommand{\Mpl}{M_\mathrm{pl}} 
\newcommand{\mpl}{m_\mathrm{pl}} 

\newcommand{\rr}{\mathrm}
\newcommand{\vev}{{\textrm{vev}}}

\newcommand{\vect}[1]{\boldsymbol{#1}}
\newcommand{\deriv}[2]{#1_{\negthinspace,#2}}

\newcommand{\calS}{\mathcal{S}}
\newcommand{\calM}{\mathcal{M}}
\newcommand{\calI}{\mathcal{I}}
\newcommand{\calH}{\mathcal{H}}
\newcommand{\calL}{\mathcal{L}}

\newcommand{\uc}{\mathrm{c}}
\newcommand{\ud}{\mathrm{d}}
\newcommand{\uB}{\mathrm{B}}
\newcommand{\ueff}{\mathrm{eff}}
\newcommand{\uini}{\mathrm{ini}}
\newcommand{\uend}{\mathrm{end}}
\newcommand{\ui}{\mathrm{i}}
\newcommand{\Veff}{V_\ueff}
\newcommand{\coupling}{\kappa}

\begin{document}

\preprint{ULB-TH/09-19}
\preprint{CP3-09-43}

\title{Fractal initial conditions and natural parameter values in
  hybrid inflation}


\author{S\'ebastien Clesse} \email{seclesse@ulb.ac.be}
\affiliation{Service de Physique Th\'eorique, Universit\'e Libre de
  Bruxelles, CP225, Boulevard du Triomphe, 1050 Brussels, Belgium}
\affiliation{Center for Particle Physics and Phenomenology, Louvain
  University, 2 chemin du cyclotron, 1348 Louvain-la-Neuve, Belgium}

\author{Christophe Ringeval} \email{ringeval@fyma.ucl.ac.be}
\affiliation{Center for Particle Physics and Phenomenology, Louvain
  University, 2 chemin du cyclotron, 1348 Louvain-la-Neuve, Belgium}

\author{Jonathan Rocher}
\email{jrocher@ulb.ac.be}
\affiliation{Service de Physique Th\'eorique, Universit\'e Libre de Bruxelles, 
CP225, Boulevard du Triomphe, 1050 Brussels, Belgium}

\date{\today}

\begin{abstract}
  We show that the initial field values required to produce inflation
  in the two fields original hybrid model, and its supergravity F-term
  extension, do not suffer from any fine-tuning problem, even when the
  fields are restricted to be sub-planckian and for almost all
  potential parameter values. This is due to the existence of an
  initial slow-roll violating evolution which has been overlooked so
  far. Due to the attractor nature of the inflationary valley, these
  trajectories end up producing enough accelerated expansion of the
  universe. By numerically solving the full non-linear dynamics, we
  show that the set of such successful initial field values is
  connected, of dimension two and possesses a fractal boundary of
  infinite length exploring the whole field space. We then perform a
  Monte--Carlo--Markov--Chain analysis of the whole parameter space
  consisting of the initial field values, field velocities and
  potential parameters. We give the marginalised posterior probability
  distributions for each of these quantities such that the universe
  inflates long enough to solve the usual cosmological
  problems. Inflation in the original hybrid model and its
  supergravity version appears to be generic and more probable by
  starting outside of the inflationary valley. Finally, the
  implication of our findings in the context of the eternal
  inflationary scenario are discussed.
\end{abstract}

\maketitle

\section{Introduction}

The paradigm of
inflation~\cite{Starobinsky:1980te,Guth:1980zm,Linde:1981mu,Inflation+25}
is currently the simplest way to solve the standard cosmological
problems and explain the Cosmic Microwave Background (CMB)
anisotropies observed so far, though other alternative mechanisms have
been proposed~(for a review see~\cite{Peter:2008qz} and references
therein). Many models of inflation have been proposed
\cite{Lyth:1998xn,Linde:2005ht}, based on single field or multi-field
potentials. If single field models are efficient effective models,
hybrid models explore the possibility that the inflaton is coupled to
other scalar fields, as first proposed by
Linde~\cite{Linde:1993cn}. When coupled to a Higgs-type field,
inflation is realized in the so-called ``inflationary valley'' when
the Higgs vacuum expectation value (\vev) vanishes and the inflation
end is triggered when the Higgs becomes tachyonic and develops and
non-vanishing vacuum expectation value (\vev).  Similar models have
rapidly been constructed in various theoretical
frameworks~\cite{Lazarides:1995vr,Jeannerot:2000sv,Kallosh:2003ux},
the most popular of them being the supersymmetric/supergravity
versions of F-term or D-term
inflation~\cite{Dvali:1994ms,Copeland:1994vg,Binetruy:1996xj,Halyo:1996pp}.

In the limit of sub-planckian field values, all hybrid inflation
models were however thought to require extremely fine-tuned initial
field values to produce enough e-folds of acceleration, from the
original model proposed by Linde to most supersymmetric
versions~\cite{Lazarides:1996rk,Tetradis:1997kp,Lazarides:1997vv},
with the exception of hilltop potentials which assume that inflation
takes place near a maximum of the potential~\cite{Boubekeur:2005zm,
  Kohri:2007gq}. The successful initial field values were found
located only in an extremely narrow band around the inflationary
valley, or on a few scattered points away from
it~\cite{Mendes:2000sq}. This was considered as a fine-tuning problem
for these models since any pre-inflationary era would have to be
fine-tuned to allow inflation to last long enough to solve the
standard cosmological problems. This fine-tuning has recently been
revisited in Ref.~\cite{Clesse:2008pf} for the original hybrid model
as well as for the supersymmetric ``smooth'' and ``shifted'' models.
Using higher precision, it was shown that the successful initial field
values are rather organised in intricate dense regions outside of the
inflationary valley (see for instance Fig.~7 of
Ref.~\cite{Clesse:2008pf}). The area occupied by these regions was
found to represent up to $15\%$ of the sub-planckian field regime for
the original hybrid model and up to $80\%$ for smooth hybrid
inflation. The physical explanation of these new successful regions
comes from the existence of an initial fast-roll phase during which
the fields roll down the potential in a chaotic way followed by a
climbing up of the valley and a slow-rolling phase back down.

However, as discussed in Ref.~\cite{Clesse:2008pf}, these new
successful regions appeared to depend on the shape of the potential,
and therefore on the potential parameters. One may wonder whether
these features are a new solution of the fine-tuning problem, i.e. if
they are robust with respect to the potential parameters. Moreover,
Ref.~\cite{Clesse:2008pf} did not discuss the statistical properties
of this space and the effect of the initial field velocities which
were assumed vanishing. Finally, the popular supersymmetric
extensions, F- or D-term hybrid models, were not studied.  The purpose
of the present paper is to quantify how these new successful
inflationary regions are widespread in the higher dimensional space of
all the model parameters, i.e. by considering not only the initial
field values but also their initial velocities and the potential
parameters.  We also extend our analysis to the F-term hybrid model,
studied in SUGRA \cite{Dvali:1994ms,Copeland:1994vg}.

In order to deal with a multi-dimensional parameter space, after
having discussed the fractal nature of the successful inflationary
regions, we introduce a probability measure and perform their
exploration by using Monte--Carlo--Markov--Chains (MCMC) methods. The
outcome of our approach is a posterior probability distribution on the
model parameters, initial velocities and field values such that
inflation lasts more than $60$ e-folds\footnote{Such probability
  distributions are almost independent of the chosen number of
  e-folds: once the field rolls down in a flat enough region of the
  potential, the total number of e-folds generated is always very
  large.}. As will be shown in the following, thanks to inflation
starting ``out of the valley'', a high number of e-folding appears to
be generic, and favoured, in the original hybrid model for parameter
ranges covering several orders of magnitude. We have also checked that
such a result is not peculiar to a given potential by applying the
same analysis to the more realistic two-field F-term inflation
potential. This treatment allows us to establish 
natural bounds on the parameters (or combination of parameters) for
each of these scenarios.

At this point, we would like to emphasize that our aim is not (yet) to
constrain these models with the current Cosmic Microwave Background
(CMB) and astrophysical data but rather to discuss in details their
ability to generate an inflationary phase. In particular, in the small
field limit, original hybrid model are known to generate a blue
spectrum of scalar initial perturbations\footnote{This conclusion can
  be altered when additional couplings are assumed for the
  inflaton~\cite{Rehman:2009wv}.}, which is disfavoured by recent CMB
experiments~\cite{Komatsu:2008hk}.  Our use of this model here is
motivated by its simplicity and its representativity of the non-linear
two-field dynamics. The more realistic F-term SUGRA model is in agreement
with the current CMB data: it predicts an almost scale invariant
spectrum and the generic formation of cosmic strings \cite{Jeannerot:2003qv}, 
a combination which was shown to be favoured by observations in 
Ref.~\cite{Bevis:2007gh}.

The paper is organised as follows. In the following section, we discuss
the fractal nature of the successful regions of inflation in the
original hybrid model and define a probability measure over the full
parameter space. In Sec.~\ref{section:mcmc}, the MCMC method is
introduced and we study step by step the effect of the initial field
velocities and the potential parameters on the probability of
obtaining $60$ e-folds of inflation. We then present the full
posterior probability distributions of these parameters for the original hybrid
scenario. In Sec.~\ref{section:fsugra}, we perform the same analysis of 
the F-SUGRA hybrid potential. Some conclusions and perspectives are 
finally presented in the last section.

\section{Fractal initial field values}

\label{sec:fractalic}

\subsection{The model}

The original hybrid model of inflation was proposed in
Refs.~\cite{Linde:1993cn,Copeland:1994vg}, its potential reads
\begin{equation} \label{eq:potenhyb2d}
V(\phi,\psi) = \frac{1}{2} m^2 \phi^2 + \frac {\lambda}{4} \left(\psi^2
- M^2 \right)^2 +\frac{\lambda'}{2} \phi^2 \psi^2.
\end{equation}
The field $\phi$ is the inflaton and $\psi$ is the auxiliary
Higgs-type field while $\lambda$, $\lambda'$ are two positive coupling
constants and $m$, $M$ are the two mass parameters. 
Inflation is assumed to be realized in the
false-vacuum along the valley\footnote{Throughout the paper
  $\langle.\rangle$ denotes the \vev~of a field.}
$\langle\psi\rangle=0$ and ends due to a tachyonic instability of
$\psi$ when the inflaton reaches a critical value $\phi_{\rr c} = M
\sqrt{\lambda / \lambda'}$. The classical system evolves toward
its true minimum $\langle\phi\rangle=0$, and $\langle\psi\rangle=\pm
M$ whereas in a realistic scenario one expects the tachyonic
instability to trigger a preheating era~\cite{Kofman:1997yn,
  Garcia-Bellido:1997wm, Felder:2000hj, Senoguz:2004vu, Micha:2004bv,
  Allahverdi:2007zz}.

To observe the effects of varying the free parameters in the dynamics of 
inflation, it is more convenient to rewrite the potential into
\begin{equation} \label{eq:potenhyb2dNEW}
V(\phi,\psi) = \Lambda^4 \left[  \left( 1 - \frac{\psi^2}{M^2} \right)^2 
+ \frac{\phi^2}{\mu^2} + \frac{\phi^2 \psi^2}{\nu^4}\right] ,
\end{equation}
under which $M,\mu,\nu$ are three mass parameters. With this
expression, the critical point of instability now reads
\begin{equation}
\label{eq:critical}
\phi_{\rr c} =   \frac{\sqrt 2 \nu^2}{M}\,.
\end{equation}
It is common usage to consider the effective one-field potential by
restricting the field dynamics to the inflationary valley and one gets
\begin{equation}
  \Veff(\phi) = \Lambda^4 \left[ 1 + \left( \frac \phi \mu \right)^2  \right].
\end{equation}   

\subsection{Equations of motion}
In a flat Friedmann--Lema\^{\i}tre-Robertson--Walker (FLRW) metric,
the equations governing the two-field dynamics are the
Friedmann-Lema\^{\i}tre equations\footnote{Throughout the paper,
  $\mpl$ denotes the physical Planck mass, and $\Mpl$ stands for the
  reduced Planck mass $\Mpl \simeq 0.2 \mpl\simeq 2.4\times
  10^{18}$~GeV.},
\begin{equation} \label{eq:FLtc12field}
\begin{split}
H^2 &= \frac {8\pi }{3 \mpl^2}  \left[ \frac 1 2 \left(\dot
\phi^2 + \dot \psi^2 \right)  + V(\phi,\psi) \right], \\
\frac{\ddot a }{a} &= \frac {8\pi}{3 \mpl^2} \left[ - \dot \phi^2
- \dot \psi^2 + V(\phi,\psi ) \right]~,
\end{split}
\end{equation}
as well as the Klein-Gordon equations 
\begin{equation} \label{eq:KGtc2field}
\begin{split}
&\ddot \phi + 3 H \dot \phi + \frac {\partial
V(\phi,\psi)}{\partial \phi} = 0~, \\
&\ddot \psi + 3 H \dot \psi + \frac {\partial
V(\phi,\psi)}{\partial \psi} = 0~.
\end{split}
\end{equation}
where $H \equiv \dot a/a$ is the
Hubble parameter, $a$ is the scale factor and a dot denotes derivative
with respect to cosmic time.

In order to study the two-fields dynamics of the hybrid model, without
assuming slow-roll, one has to integrate these equations numerically
from a given set of initial conditions (IC) for the field
values. Throughout the paper we will define a successful IC as a point
in field space that lead to a sufficiently long phase of inflation to
solve the horizon and flatness problem. We will assume that
$N=\ln(a/a_\uini)\simeq 60$ e-folds is the critical value required,
though this value can change by a factor of two depending on the
reheating temperature and the Hubble parameter at the end of
inflation~\cite{Liddle:2003as, Ringeval:2007am}. However, generically,
once inflation starts it lasts for much more than $60$ e-folds and our
results are not sensitive to the peculiar value chosen.

\subsection{The set of successful initial field values}

\begin{figure}[]
  \includegraphics[width=8.7cm]{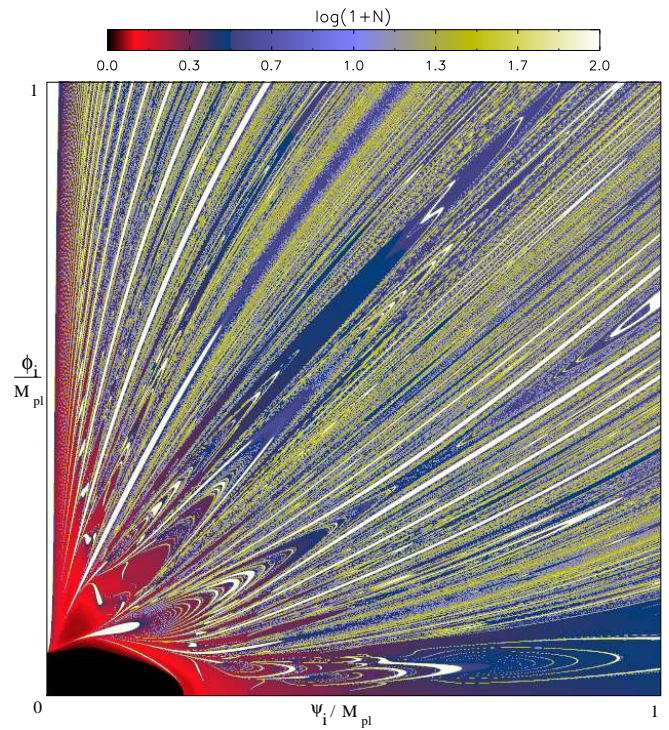}
  \caption{Mean number of e-folds obtained from $512^2$ initial
    field values in the plane $(\psi_\ui/\Mpl,\phi_\ui/\Mpl)$. This
    figure has been obtained by averaging the number of e-folds
    (truncated at $100$) produced by $2048^2$ trajectories down to
    $512^2$ pixels. The potential parameters have been set to
    $M=0.03 \mpl$, $\mu = 636 \mpl$, $\nu^2 = 3 \times
    10^{-4}\mpl^2$.}
  \label{fig:grid}
\end{figure}

As already mentioned in the introduction, the space of successful IC
for the field values alone has been discussed in
Ref.~\cite{Clesse:2008pf} and found to be composed of a intricate
ensemble of points organized into continuous patterns. In
Fig.~\ref{fig:grid}, we have represented the mean number of e-folds
generated at each sub-Planckian initial field values, for a set of
\emph{fixed} potential parameters and assuming vanishing initial
velocities. We have computed the trajectories obtained from
$2048^2$ initial field values and stopped the integration when the
fields are trapped in one of the minimum of the potential, i.e. for
$H^2 \le V/(3\Mpl^2)$, or when the accelerated expansion exceeds
$10^2$ efolds. The resulting grid has a small intricated structure of
successful regions spread over the whole plane which ends up being
difficult to represent in a figure. As a result, we have chosen to
present in Fig.~\ref{fig:grid} a downgraded $512^2$ pixels image in
which each pixel has been given a color according to the average
number of e-folds obtained in our original $2048^2$ grid. A given
pixel may therefore hide both successful and unsuccessful initial
field values and the color measures their relative density. An higher
resolution image would be self-similar to Fig.~\ref{fig:grid}, with
more thinner successful domains visible.

Notice that we recover the inflationary valley as the white
vertical narrow strip located along $\psi_\ui=0$ whereas the minima of
the potential are along the horizontal axis at $\psi=\pm 0.15\Mpl$
(for $M=0.03\mpl$ as chosen in the figure). The black region in
Fig.~\ref{fig:grid} precisely corresponds to the trajectories
``below'' the critical point $\phi<\phi_\uc$ which are fast-rolling
inside the minima. In analogy with the anamorphosis of light
produced by a distorted mirror, each point outside the inflationary
valley is connected by a trajectory to a point inside the inflationary
valley. The trajectory first fast-rolls towards the bottom of the
potential, and after a few rebounds becomes oriented along the valley,
climbs it and then produces inflation when slow-rolling back
down. There is thus a one-to-one correspondence between the IC and the
point in the valley for which the trajectory stops to climb and starts
to slow-roll.

It was shown in Ref.~\cite{Clesse:2008pf} that such ``anamorphosis
points'' can cover up to $15\%$ of the total area when restricting the
IC to sub-planckian values.  Moreover, as can be checked in
Fig.~\ref{fig:grid}, these regions exhibits a fractal looking
aspect. Before studying the influence of the potential parameters and
initial field velocities, one may wonder if the area of this
two-dimensional set of points is indeed well-defined? Equivalently, do
new successful regions appear inside unsuccessful domains and
conversely? In order to quantify how much the anamorphosis points are
a probable way to have inflation in the whole parameter space, we
first address the question of defining a measure on the initial field
values space. In particular, this requires to determine the dimension
of the set
\begin{equation}
\label{eq:Sdef}
\calS \equiv \left\{(\phi_\ui, \psi_\ui) \ \diagup \ N>60 \right\}.
\end{equation}

\subsection{Chaotic dynamical system}

\subsubsection{Phase space analysis}

As suggested by Fig.~\ref{fig:grid}, at fixed potential parameter
values, the dynamical system defined by Eqs.~(\ref{eq:FLtc12field})
and (\ref{eq:KGtc2field}) seems to exhibit a chaotic behaviour. In
particular, the sensitivity of the trajectories to the initial field
values comes from the presence of three attractors. Two of them are
the global minima of the potential, $\calM_\pm$ respectively at
$(\phi=0,\psi = \pm M)$, in which all classical trajectories will end,
whereas the less obvious is a quasi-attractor $\calI$ defined by the
inflationary valley itself ($\psi=0$, $\phi>\phi_\uc$). Indeed,
whatever the initial field values, as soon as the system enters
slow-roll one has (in Planck units)~\cite{Ringeval:2005yn},
\begin{equation}
v^2 \equiv \left(\dfrac{\ud \phi}{\ud N}\right)^2 + \left(\dfrac{\ud \psi}{\ud N}
\right)^2 = 2 \epsilon_1 \ll 1,
\end{equation}
where $\epsilon_1$ is the first Hubble flow
function~\cite{Schwarz:2001vv}. The system therefore spends an
exponentially long amount of cosmic time in the valley. The
sensitivity to the initial conditions comes from the presence of these
three attractors: either the trajectory ends rapidly into one of the
two minima, or it lands on the valley where it freezes.

\begin{figure}[]
  \includegraphics[width=8cm]{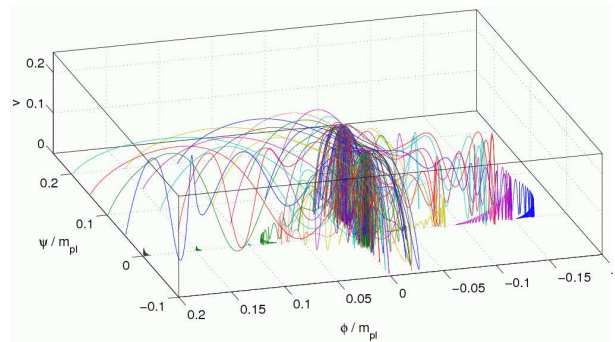}
  \caption{Phase space $v^2(\phi, \psi)$ for 25 trajectories and
    vanishing initial velocities. The potential parameters are fixed
    to the values $M=0.03 \mpl, \mu = 636 \mpl , \nu = 6.36 \times
    10^{-4}$. All trajectories end on the three attractor of the
    dynamical system: the two global minima of the potential, and the
    inflationary valley with almost vanishing slow-roll
    velocity. These three attractors induce the chaotic behavior.}
  \label{fig:phasespace}
\end{figure}

A phase space plot is represented in Fig.~\ref{fig:phasespace} in
which we have computed $25$ trajectories from a grid of initial field
values. The inflationary valley clearly appears as the attractor with
quasi null velocity vector ($\epsilon_1 \ll 1$), while around the two
global minima, two ``towers'' appear due to the field oscillations
around them.

\subsubsection{Basins of attraction}

From the definition of $\cal S$ in Eq.~(\ref{eq:Sdef}), one has
\begin{equation}
\calS = F^{-1}(\calI),
\end{equation}
where $F(\phi,\psi)$ stands for the mapping induced by the
differential system of equations (\ref{eq:FLtc12field}) and
(\ref{eq:KGtc2field}). The set of successful initial field values
$\calS$ is therefore the basin of attraction of the attractor
$\calI$~\cite{Ott:fractals, Falconer:fracgeo}. Since the attractor
$\calI$ is a dense set of dimension 2 and $F$ is continuous, one
expects $\calS$ to contain a dense set of dimension
2~\cite{Falconer:fracgeo}. As can be intuitively guessed, the boundary
of $\calS$ can however be of intricate structure because of the
sensitivity to the initial conditions: two trajectories infinitely
close initially can evolve completely differently. As we show in the
following, $\calS$ is actually a set of dimension two having a fractal
boundary of dimension greater than one.

Finally, by the definition of a continuous mapping, all parts of
$\calS$, boundary included, must be connected together and to the
inflationary valley $\calI$. The fractal looking aspect of
Fig.~\ref{fig:grid} is only induced by the intricate boundary
structure of $\calS$ which is exploring all the initial field values
space. The fractality of the boundaries of the space of initial fields
values was first mentioned in Ref.~\cite{Ramos:2001zw}, but the study
was restricted to a small region of the field space and the model
included dissipative coefficients. As an aside remark, let us notice
that the existence of a fractal boundary may have strong implications
in the context of eternal chaotic inflation: there would exist
inflationary solution close to any initial field values.

In order to quantify the chaotic properties of the dynamical
system defined by the mapping $F(\phi,\psi)$, we turn to the
calculation of the Lyapunov exponents.

\subsubsection{Lyapunov exponents}

\begin{figure}[]
  \includegraphics[width=8.7cm]{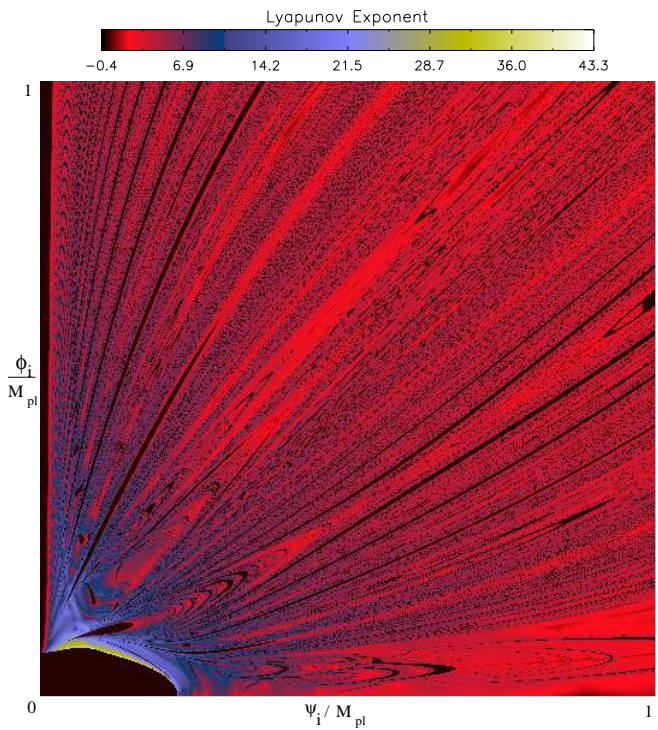}
  \caption{Highest Lyapunov exponent as a function of the initial
    field values in the orginal hybrid model. The potential parameters
    are the same as in Fig.~\ref{fig:grid}. The field evolution is
    therefore stable on the set $\calS$ of successful initial field
    values (black) but exhibits chaotic behaviour elsewhere.}
  \label{fig:lyap}
\end{figure}

The Lyapunov exponents at an initial point
$\vect{\chi}_\ui=(\phi,\psi,\deriv{\phi}{N},\deriv{\psi}{N})|_\ui$
measures how fast two infinitely close trajectories mutually diverge
or converge. They give a mean to characterize the stretching and
contracting characteristics of sets under the mapping induced by the
differential system. A small perturbation $\vect{\delta \chi}$ around
the trajectory $\vect{\chi}(N)$ will evolve according to
\begin{equation}
\label{eq:pertevol}
\dfrac{\ud \vect{\delta \chi}}{\ud N} = \ud F \cdot \vect{\delta \chi}\,,
\end{equation}
where $\ud F$ stands for the Jacobian of the differential system
$F$. The Lyapunov exponents at the initial point $\vect{\chi}_\ui$ and
along the direction $\vect{\delta \chi}_0$ are the numbers defined
by~\cite{Ott:fractals}
\begin{equation}
  h(\vect{\chi}_\ui,\vect{\delta \chi}_0) = \lim_{N\to\infty} \dfrac{1}{N}\ln \dfrac{
    \left|\vect{\delta \chi}(N)\right|}{\left|\vect{\delta \chi}_0 \right|}\,,
\end{equation}
where $\vect{\delta \chi}(N)$ is the solution of
Eq.~(\ref{eq:pertevol}) with $\vect{\delta \chi}(0)=\vect{\delta
  \chi}_0$ and $\vect{\chi}(0)=\vect{\chi}_\ui$. If the considered set
is an attractor or an invariant set of the differential system having
a natural measure, one can show that the exponents do not depend on
the initial point $\vect{\chi}_\ui$. At fixed potential parameters,
there are four Lyapunov exponents associated with the differential
system of Eqs.~(\ref{eq:FLtc12field}) and (\ref{eq:KGtc2field}). If
the largest exponent is positive, then the invariant set is chaotic.

In Fig.~\ref{fig:lyap}, we have computed the largest Lyapunov exponent
at each point of the plane $(\phi_\ui,\psi_\ui)$. The numerical method
we used is based on Refs.~\cite{1997:Dieci, 2003:dieci} and uses the
public code $\texttt{LESNLS}$. Let us notice that since $\calI$ is
only a quasi-attractor, we have stopped the evolution at most when
$H_\uend^2=V/(3\Mpl^2)$, i.e. just before the fields would classically
enter either $\calM_+$ or $\calM_-$. As can be seen, all points
belonging to $\calS$ exhibit the same and small negative Lyapunov
exponent: the invariant set $\calS$ is therefore non-chaotic. On the
other hand, all the other initial field values associated with the
basins of attraction of $\calM_\pm$ have a positive Lyapunov
exponent. For those, the field evolution is chaotic and exhibits a
sensitivity to the initial conditions. Notice that these exponents may
slightly vary from point to point due to our choice to stop the
integration at $H_\uend$ instead of the classical attractors
$\calM_\pm$. This is particularly visible for the trajectories
starting close to $H_\uend$ (green shading): there is not enough
evolution to get ride of the transient evolution associated with the
initial conditions.

\subsection{Fractal dimensions of $\calS$ and its boundary}

\subsubsection{Hausdorff and box-counting dimension}\label{sec:boxcountingdim}

Since we suspect a set with fractal properties, the natural
measure over $\calS$, extending the usual Lebesgue measure, is the
Hausdorff measure. The $s$-dimensional Hausdorff measure of $\calS$ is
defined by~\cite{Falconer:fracgeo}
\begin{equation}
\calH^s(\calS) = \lim_{\delta \to 0} \inf \left\{ \sum_{i=1}^\infty
      |U_i|^s\ \diagup \ \calS \subset \bigcup_{i=1}^\infty
      U_i\ ; \ |U_i| \le \delta \right\}.
\end{equation}
In this definition, the sets $U_i$ form a $\delta$-covering of $\calS$
and the diameter function has been defined by $|U|\equiv \sup\{|x-y|\
\diagup \ x,y \in U\}$. As a result, $\calH^s(\calS)$ is the smallest
sum of the $s$th powers of all the possible diameters $\delta$ of all
sets covering $\calS$, when $\delta \to 0$. Having such a measure, the
fractal dimension of $\calS$ is defined to be the minimal value of $s$
such that the Hausdorff measure remains null (or equivalently the
maximal value of $s$ such that the measure is infinite). In practice,
measuring the Hausdorff dimension using this definition is not
trivial, due to the necessity of exploring all
$\delta$-coverings. However, in our case, we are interested in the
fractal properties of a basin of attraction associated with a
continuous dynamical system and one can instead consider the so-called
box-counting dimension~\cite{Falconer:fracgeo}. This method simply
restricts the class of the $U_i$ to a peculiar one, all having the
same diameter $\delta$. When the mapping $F$ is self-similar, one can
show that box-counting and Hausdorff dimensions are equal. In general,
the Hausdorff dimension is less or equal than the box-counting
one. Here, $F$ being a contracting continuous flow, we expect the
equality to also hold.

To define the box-counting dimension, we cover the set $\calS$ with
grids of step size $\delta$, and count the minimal number of boxes
$N(\delta)$ necessary for the covering. The box-counting dimension is
then given by
\begin{equation}\label{eq:boxcountingdim}
D_\uB= \lim _{\delta \rightarrow 0} \frac{\log N(\delta)}{\log (1/ \delta) }\,.
\end{equation}
This method has the advantage to be easily implemented numerically 
and, in the following, we will apply it to calculate the dimension of
$\calS$ and its boundary. 

\subsubsection{Fractal boundary of $\calS$}

For each randomly chosen point of the plane $(\phi_\ui,\psi_\ui)$, we compute
three trajectories. The first one starts from the point under
consideration while the two others have initial conditions modified by
$+\delta$ and $-\delta$ along one direction (for example along $\phi$,
but the chosen direction does not affect the result). For each of
these trajectories, we determine in which attractor ($\calM_\pm$ or
$\calI$) the flow ends. Since we are interested in the boundary of
$\calS$, we calculate the proportion $f(\delta)$ of points for which
at least one trajectory ends in $\calI$, and another in $\cal M_+$ or 
$\cal M_-$. The process is iterated for increasingly smaller values of 
$\delta$ and we evaluate how the area of the
$\delta$-grid covering of $\calS$ scales with $\delta$. So strictly
speaking, our evaluation of the box-counting dimension is made through
the determination of the Minkowski dimension of the boundary of
$\calS$~\cite{Falconer:fracgeo}. From Eq.~(\ref{eq:boxcountingdim}),
assuming that, at small $\delta$, 
\begin{equation}
f(\delta) \propto \delta^\alpha,
\end{equation}
the box-counting dimension of the $\calS$ boundary is then given
by~\cite{Ott:fractals}
\begin{equation}
D_\uB = 2 - \alpha.
\end{equation}

\begin{figure}[]
\includegraphics[width=8cm]{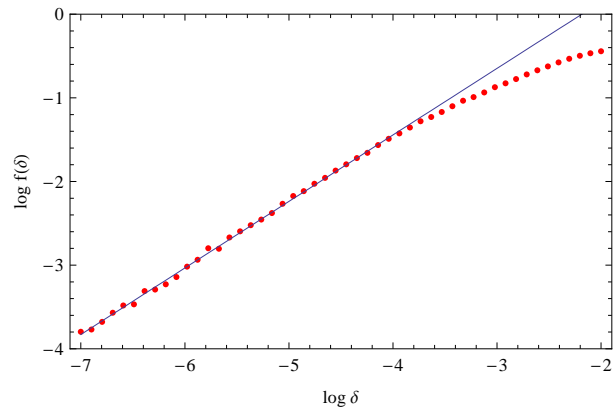}
\caption{Fraction of initial field values in a $\delta$-sized box
  intercepting the set $\calS$, as a function of $\delta$. The field
  has been restricted to sub-planckian values and the potential
  parameters are fixed to $\lambda = \lambda'=1$, $m=10^{-6} \mpl$ and
  $M=0.03 \mpl$. The exponent $\alpha$ of the power law dependency
  gives the box-counting dimension $D_\uB = 2 - \alpha \simeq 1.2$
  showing that $\calS$ possesses a fractal boundary.}
 \label{fig:dim_fractale}
\end{figure}

In Fig.~\ref{fig:dim_fractale}, we have plotted $f(\delta)$ as a
function of $\delta$ at fixed potential parameters. We recover the
expected power law behaviour, the slope of which is approximately
$\alpha \simeq 0.80$. As a result, the boundary of $\calS$ is indeed a
fractal of box-counting dimension
\begin{equation}
D_\uB \simeq 1.20.
\end{equation}
Notice that this value depends on the chosen set of potential
parameters, as one may expect since they affect the shape of $\calS$
and the typical size of the structures.

\subsubsection{Dimension of $\calS$}

\begin{figure}[]
\includegraphics[width=8cm]{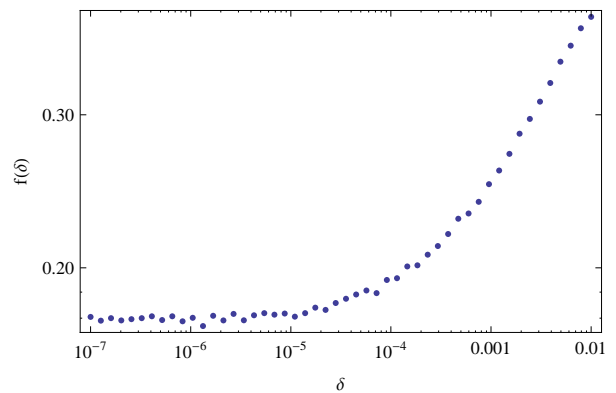}
\caption{Fraction of initial field values leading to inflation in a
  $\delta$-sized box as a function of $\delta$. The potential
  parameters are the same as in Fig.~\ref{fig:dim_fractale}. Once the
  box is small enough to be fully contained in $\calS$, $f(\delta)$
  remains constant. As a result, the box-counting dimension of $\calS$
  is $D_{\uB}=2$ and the interior of $\calS$ is not fractal.}
\label{fig:fractal_area}
\end{figure}

In order to determine the box-counting dimension of $\calS$ itself one
can apply a similar method than the one used for its boundary. Now
$f(\delta)$ denotes the proportion of points for which at least one of
the three trajectories end in the attractor $\calI$ (this condition
therefore includes also the points belonging to the boundaries). The
resulting power-law is represented in Fig.~\ref{fig:fractal_area}.

For small enough values of $\delta$, the $\delta$-sized boxes are
small enough to be fully contained in $\calS$ and the function
$f(\delta)$ appears to be constant in that case. As a result, the
box-counting dimension of $\calS$ is $2$. We therefore conclude that,
like for the well-known Mandelbrot set~\cite{Mandelbrot:1980}, the 
boundary of $\calS$ is fractal but
the set of successful inflationary points is not and has the dimension 
of a surface. Consequently, although the boundary of $\calS$
has an infinite length ($D_\uB = 1.2$), it has a vanishing area: the
Hausdorff dimension of $\calS$ (boundary included) is therefore also
2. As a result, the two-dimensional Hausdorff measure on $\calS$
reduces to the usual two-dimensional Lebesgue measure and this will
be our choice for defining a probability measure in the rest of the 
paper.

As previously
emphasized, the potential parameters and initial field velocities have
been fixed in this section and the set $\calS$ is actually the
two-dimensional section of an higher dimensional set, whose boundary
is also certainly fractal (and therefore of null measure). Since one 
can no longer use griding method to
explore such a high dimensional space, we move on in the next
section to a MCMC exploration of the full parameter space to assess
the overall probability of getting inflation in the hybrid model.

\section{Probability distributions in hybrid inflation}

\label{section:mcmc}

The aim of this section is to use Monte-Carlo-Markov-Chains (MCMC)
techniques in order to explore the whole parameter space, including the
initial field velocities and all the potential parameters. With
unlimited computing resources, we could have used a griding method to
localise the hypervolumes in which inflation occurs, as we have done
for the two-dimensional plane $(\phi_\ui,\psi_\ui)$ in the previous
section. For the original hybrid model, we have in total seven
parameters that determine a unique trajectory: two initial field
values, initial field velocities and the three potential parameters
$M$, $\mu$ and $\nu$. To probe this seven-dimensional space, more than
just measuring the hypervolume of the successful inflationary regions,
we define a probability measure over the full parameter space. Using
Bayesian inference, one can assess the posterior probability
distribution of all the parameters to get enough e-folds of
inflation. Monte--Carlo--Markov--Chains (MCMC) method is a widespread
technique to estimate these probabilities, its main power being that
it numerically scales linearly with the number of dimensions, instead 
of exponentially.

Several algorithms exist in order to construct the points of a Markov
chain, the Metropolis--Hastings algorithm being probably the
simplest~\cite{Metropolis53,Hastings70}. Each point $x_{i+1}$,
obtained from a Gaussian random distribution (the so-called proposal
density) around the previous point $x_i$, is accepted to be the next
element of the chain with the probability
\begin{equation}
\label{eq:mcmctrans}
P(x_{i+1}) = \min \left[1,\frac{\pi(x_{i+1})}{\pi(x_i)}  \right],
\end{equation}
where $\pi(x)$ is the function that has to be sampled via the Markov
chain. MCMC methods have been intensively used in the context of CMB
data analysis~\cite{Christensen:2001gj, Lewis:2002ah, Martin:2006rs,
  Lorenz:2007ze, Dunkley:2008ie} where the function $\pi(\theta|d)
\propto \calL(d|\theta)P(\theta)$ is the posterior probability
distribution of the model parameters given the data. In the context of
Bayesian inference, this one is evaluated from the prior distributions
$P(\theta)$ and the likelihood of the experiment
$\calL(d|\theta)$. After a relaxation period, one can show that
Eq.~(\ref{eq:mcmctrans}) ensures that $\pi$ is the asymptotic
stationary distribution of the chain~\cite{MacKayBook}. The MCMC
elements directly sample the posterior probability distribution
$\pi(\theta|d)$ of the model given the data.

In our case, we can similarly define a likelihood $\calL$ as a
binary function of the potential parameters, initial field values and
velocities. Either the trajectory ends up on $\calI$ and produces more 
than $60$ e-folds of inflation, or it does not. In
the former case we set $\calL=1$ whereas $\calL=0$ for no
inflation. The function $\pi$ we sample is then defined by $\pi =
\calL P(\theta)$ where $\theta$ stands for field values, velocities
and potential parameters and $P$ is our prior probability distribution
that we will discuss in the next section.

\subsection{Prior choices}
\label{sec:priors}
MCMC methods require a prior assumption on the probability
distributions of the fields, velocities and potential parameters.  As
we only consider in this work the initial conditions and parameters
space leading to at least $60$ e-folds of inflation, the prior choices
are only based on theoretical arguments. These arguments can be linked
to the framework from which the potential is deducted. If one
considers the hybrid model to be embedded in supergravity, the 
fields have to be restricted to values less than the reduced Planck
mass. We adopt here this restriction for initial field values, not only 
because of this argument, but also because it has been shown in
Ref.~\cite{Clesse:2008pf} that if super-plankian fields are allowed,
trajectories become generically successful. On the other hand, the
model was considered to suffer some fine-tuning when one of the fields
has to be order of magnitudes smaller than the other. As inflation is
not possible for very small initial values of both fields (because of
the Higgs instability), we have considered a flat prior for initial
field values in $[-\Mpl,\Mpl]$ as opposed to a flat prior for the
logarithm of the fields. Note that one has to include
negative values of the fields in order to take into account the 
orientation of the initial velocity vector.

Concerning the initial field velocities, from the equations of motion,
one can easily show that there exist a natural limit\footnote{This is
  just the limit $\epsilon_1 < 3$ in Planck units \cite{Ringeval:2005yn}.}
\begin{equation}
  v^2 =  \left(  \frac{\dd \phi} {\dd N} \right)^2 + \left( \frac{\dd
      \psi} {\dd N} \right)^2 < 6.
\end{equation}
Similarly, our prior choices are flat distributions inside such a
circle in the plane $(\deriv{\phi}{N},\deriv{\psi}{N})$, where ``$,N$''
denotes a partial derivative with respect to the number of e-folds.

In the absence of a precise theoretical setup determining the
potential there are no a priori theoretical constraint on its
parameters $M$, $\mu$ and $\nu$. Let us just mention that for $\mu <
0.3$, the dynamics of inflation in the valley is possibly strongly
affected by slow-roll violations~\cite{Clesse:2008pf}. As a result,
with the concern to not support a particular mass scale, we have
chosen the following flat priors on the logarithm of the parameters:
\begin{equation}
\begin{aligned}
  -1 & < \log\dfrac{\mu}{\mpl} < 4, \\
 -3 & < \log\dfrac{M}{\mpl} < -0.7, \\
 -6 &< \log\dfrac{\nu^2}{\mpl^2} < 2,
\end{aligned}
\end{equation}
in which the upper and lower limits have been set for numerical
convenience, and $M\le \Mpl$.

Notice that the $\Lambda$ dependencies are not important here because
this parameter only rescales the potential and thus does not change
the dynamics.

In the next sections, we perform the MCMC exploration of the parameter
space from these priors. Firstly by reproducing the results of
Sec.~\ref{sec:fractalic} in the two-dimensional section
$(\phi_\ui,\psi_\ui)$, then by including the initial field velocities
and finally by considering all the model parameters. Unless
otherwise mentioned, the chains contain $10^6$ points, which
corresponds to one percent error on the marginalised probability
distributions. In the figures, the overall values of the posterior
probability density distributions have not been represented since they
are determined by the imposing the integral over the parameters to be
equal to one.

\subsection{MCMC on initial field values}

\begin{figure}
\includegraphics[width=8cm]{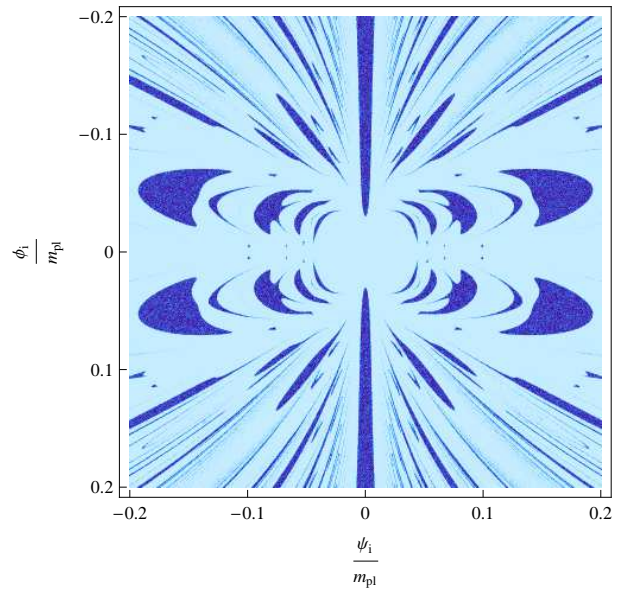}
\caption{Two-dimensional posterior probability distribution in the
  plane $(\phi_\ui,\psi_\ui)$ leading to more than $60$ e-fold of
  inflation in the hybrid model. Notice that its integral over the
  plane is normalised to unity. The dark blue regions corresponds to a
  maximal probability density whereas it vanishes elsewhere. The
  potential parameters are set to $M=0.03 \ \mpl, \nu^2 = 6.36 \times
  10^{-4}\ \mpl^2, \mu=636 \ \mpl$. As expected, the MCMC exploration
  matches with the griding methods (see Fig.~\ref{fig:grid}).}
\label{fig:2Dfield_nov}
\end{figure}

In order to test our MCMC, we have first explored the space of initial
field values leading or not to more than $60$ e-folds of
inflation. The potential parameters have been fixed to various values
already explored by griding methods in Sec.~\ref{sec:fractalic} and
Ref.~\cite{Clesse:2008pf}, while the initial velocities are still
assumed to vanish. The MCMC chain samples have been plotted in
Fig.~\ref{fig:2Dfield_nov}. Notice that to recover the fractal
structure of the boundary of $\calS$, one has to adjust the choice of
the Gaussian widths of the proposal density distribution. If those are
too large, the acceptance rate will be small because the algorithm
tends to test points far away from the last successful point, and if
they are too small the chains remain stuck in the fractal structures
without exploring the entire space. Nevertheless, with an intermediate
choice, Fig.~\ref{fig:2Dfield_nov} shows that the intricate structure
of the boundary of $\calS$ can be probed with the MCMC. More than
being just an efficient exploration method compared to griding, the
MCMC also provides the marginalised probability distributions of
$\phi_\ui$ and $\psi_\ui$ such that one gets inflation. They have been
plotted in Fig.~\ref{fig:1Dfields} (top two plots), the normalisation
being such that their integral is unity. As one can guess from
Fig.~\ref{fig:2Dfield_nov}, with vanishing initial velocities and a
fixed set of potential parameters, inflation starting in the valley is
\emph{not} the preferred case since the area under the distribution of
$\psi_\ui$ outside of the valley is larger than inside. Moreover,
these distributions take non-vanishing values everywhere and there is
therefore no fine-tuning problem. Of course, one still have to
consider the other parameters and this is the topic of the next
sections.

\subsection{MCMC on initial field values and velocities}

\begin{figure}[]
  \includegraphics[width=8cm]{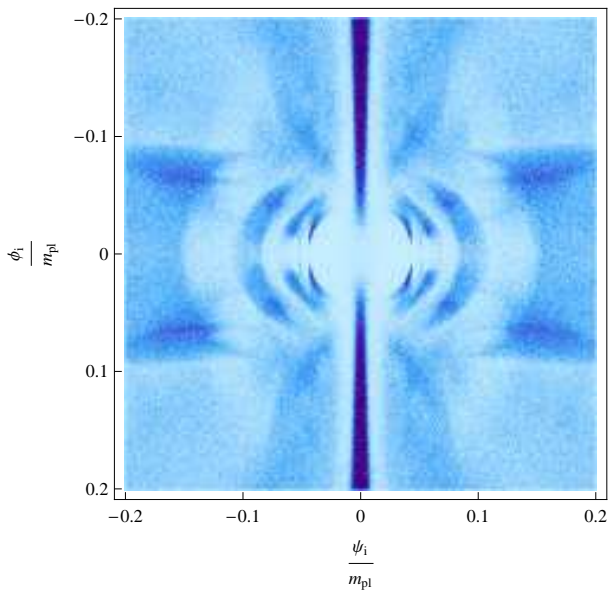}
  \caption{Two-dimensional marginalised posterior probability
    distribution for the initial fields values. The marginalisation is
    over the initial field velocities whereas the potential parameters
    are still fixed. The shading is proportional to the probability
    density value while the two-dimensional integral over the plane is
    equal to one. Although the inflationary valley has the highest
    probability density, its area remains restricted such that the
    most probable initial field values to get inflation are still out
    of the valley (see Fig.~\ref{fig:1Dfields}).}
\label{fig:2Dfield_v}
\end{figure}

The initial
values of the field velocity are inside a disk of radius $\sqrt {6}$
in the plane $(\deriv{\phi}{N},\deriv{\psi}{N})$ (in Planck
units). The marginalised two-dimensional posteriors for the initial
field values is plotted in Fig.~\ref{fig:2Dfield_v} whereas the
marginalised posterior for each field are represented in
Fig.~\ref{fig:1Dfields} (middle line). Even if non-vanishing velocities are
considered, the successful inflationary patterns remain. Notice that
they appear to be blurred simply because of the weighting induced by
marginalising the full probability distribution over the initial
velocities.

\begin{figure}[]
\includegraphics[width=6.cm]{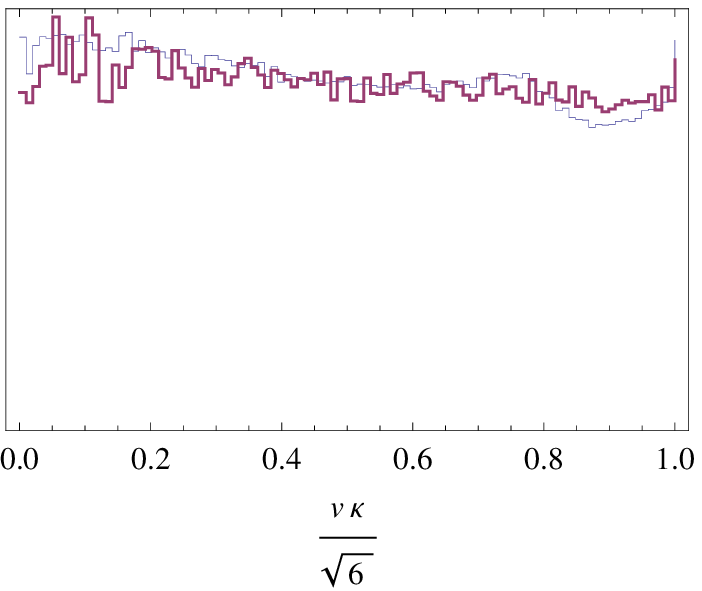}
\includegraphics[width=6.cm]{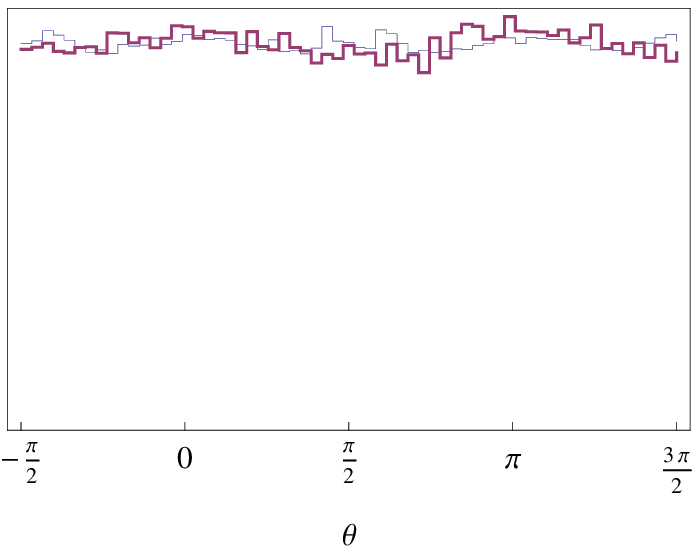}
\caption{Marginalised posterior probability distributions for the
  modulus (top) and angle (bottom) of initial field velocity. The thin
  superimposed blue (lighter) curves are obtained at fixed potential parameters, while the
  thick red are after a full marginalisation over all the model
  parameters. As expected from Hubble damping, all values are
  equiprobable since the field do not keep memory of the initial
  velocity.}
\label{fig:1Dspeeds}
\end{figure}

In Fig.~\ref{fig:1Dspeeds}, we have also represented the marginalised
posterior probability distribution for the modulus and direction of
the initial velocity vector. Their flatness implies that there are no
preferred values.  This is an important result because one could think
that large initial velocities could provide a way to kick trajectories
in or out of the successful regions.  This actually never happens
because of the Hubble damping term in the Friedmann equations, allowing 
only a generation of a small number of
e-folds before the trajectory falls in one of the three attractors.

\subsection{MCMC on initial field values, velocities and potential
  parameters}

\begin{figure}[]
  \includegraphics[width=8cm]{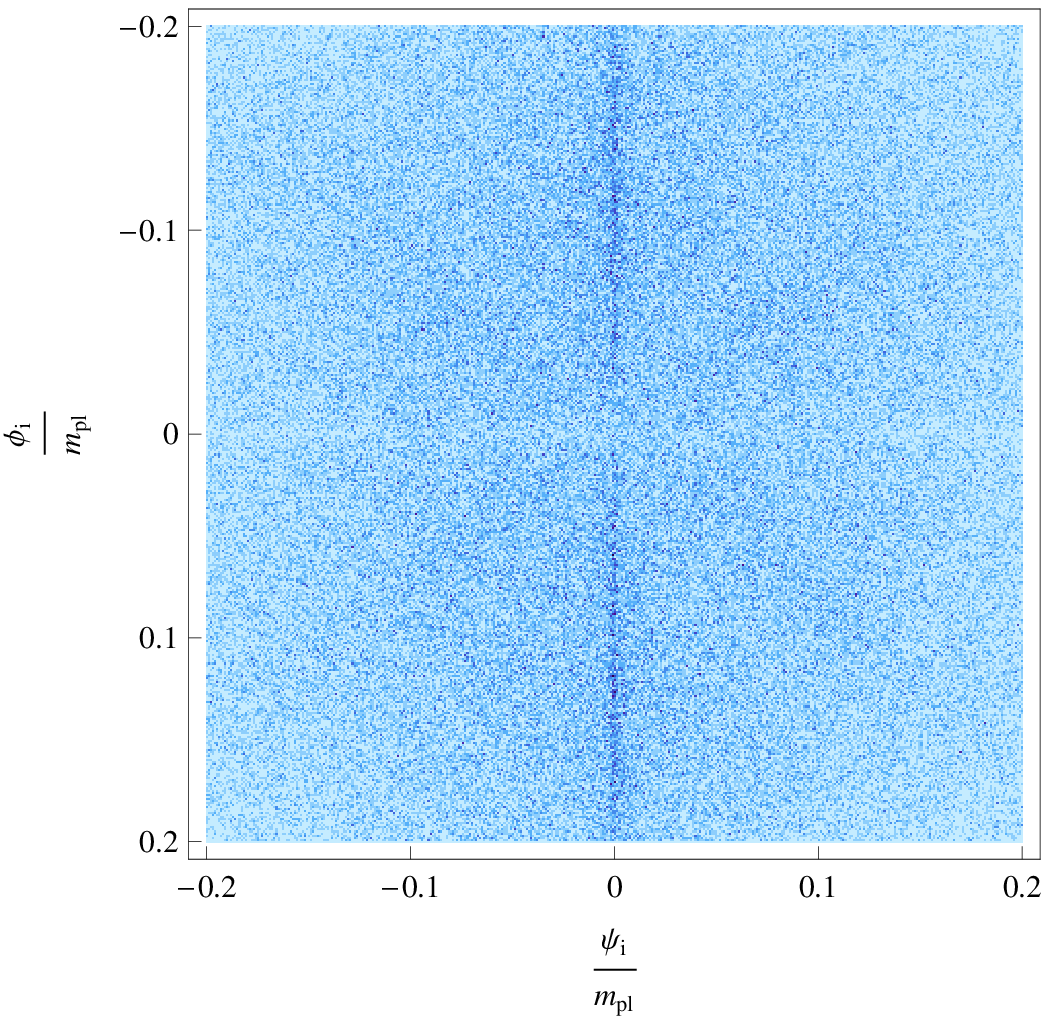}
  \caption{Two-dimensional marginalised posterior probability
    distribution for the initial fields values. The marginalisation is
    over the initial field velocities and all the potential
    parameters. The shading is proportional to the probability density
    value. The inflationary valley is still visible around
    $\psi_\ui=0$ and the posterior takes non-vanishing values
    everywhere in the $(\phi_\ui,\psi_\ui)$ plane.}
\label{fig:2Dfield_all}
\end{figure}

\begin{figure*}[]
\includegraphics[width=14cm]{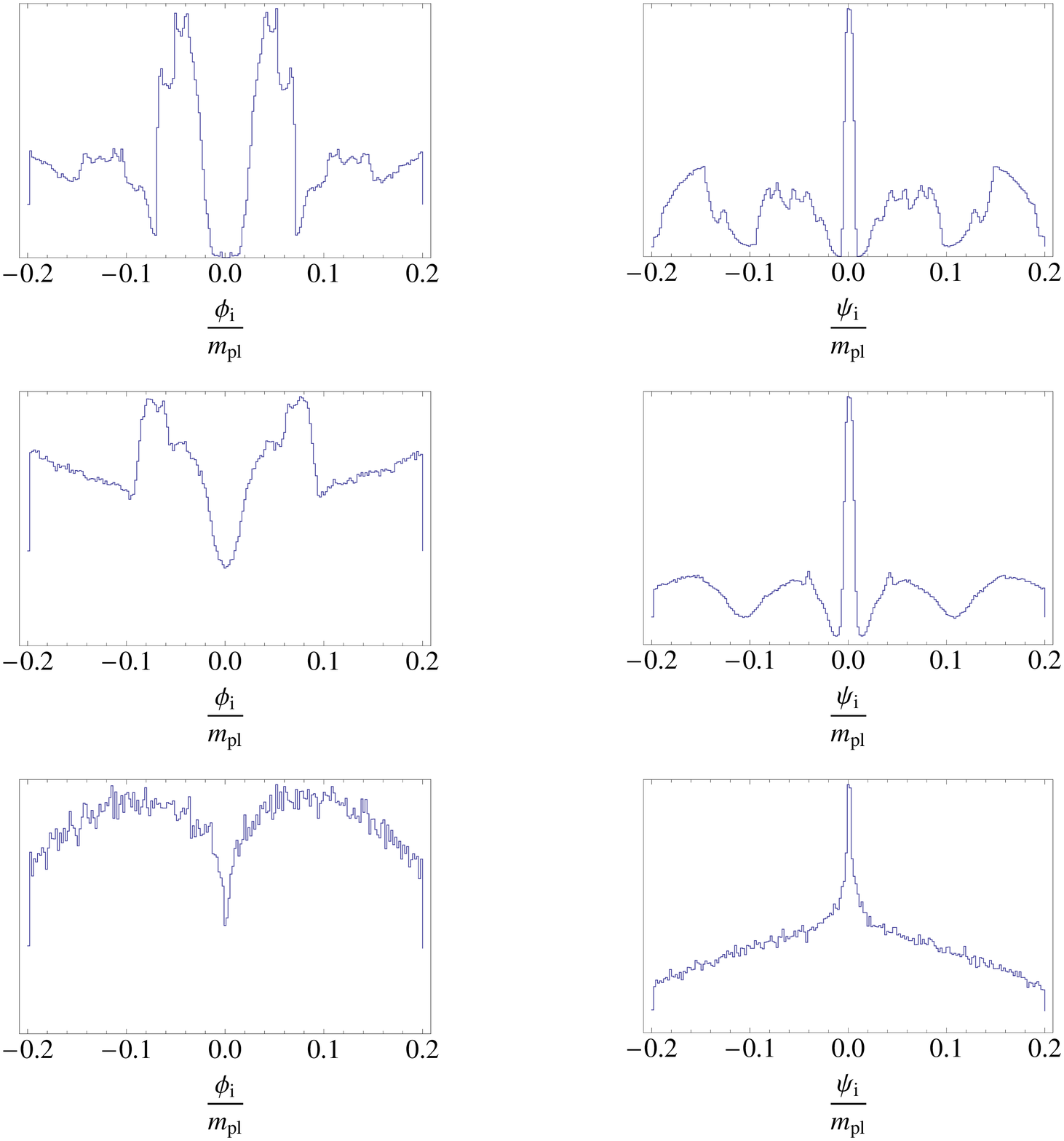}
\caption{Marginalised posterior probability distributions for the
  initial field values $\phi_\ui$ and $\psi_\ui$. The top panels
  correspond to vanishing initial velocities and fixed potential
  parameters, the middle ones are marginalised over velocities at
  fixed potential parameters, while the lower panels are marginalised
  over velocities and all the potential parameters.}
\label{fig:1Dfields}
\end{figure*}

The most interesting part of the exploration by MCMC technique
concerns the study of the full parameter space. The only restriction
being associated to the necessity of $M< \Mpl $ as discussed in
Sec.~\ref{sec:priors}. The chains contain $200000$ elements and the
estimated error on the posteriors is about a few percents.

We have plotted in Fig.~\ref{fig:2Dfield_all} the marginalised
two-dimensional posterior for the initial field values. In comparison
with Fig.~\ref{fig:2Dfield_nov} and \ref{fig:2Dfield_v}, the most
probable initial field values are now widespread all over the
accessible values; the intricate patterns that were associated with
the successful field values (at fixed potential parameters) are now
diluted over the full parameter space. The resulting one-dimensional
probability distributions for each field are plotted in
Fig.~\ref{fig:1Dfields} (bottom panels). One can observe that the
$\psi$ distribution is nearly flat outside the valley but remains
peaked around a extremely small region around $\psi=0$. Integrating
over the field values, initial conditions outside the valley are still
the preferred case.

Concerning the probability distributions of the modulus $v$ and the
angular direction $\theta$ of the initial velocity vector, results integrated
over the whole parameter space do not present qualitative differences
compared to the posteriors with fixed potential parameters, as one may
expect since the Hubble damping prevents the initial velocities to
influence the dynamics (see Fig.~\ref{fig:1Dspeeds}).

\begin{figure}
\includegraphics[width=6cm]{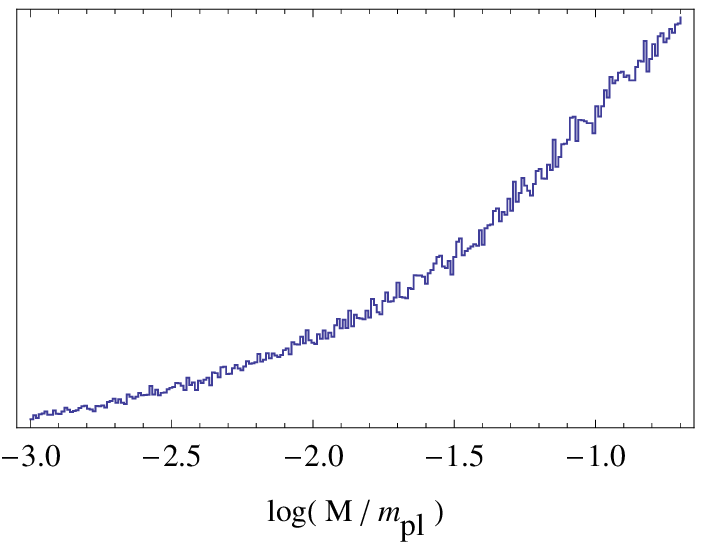}
\includegraphics[width=6cm]{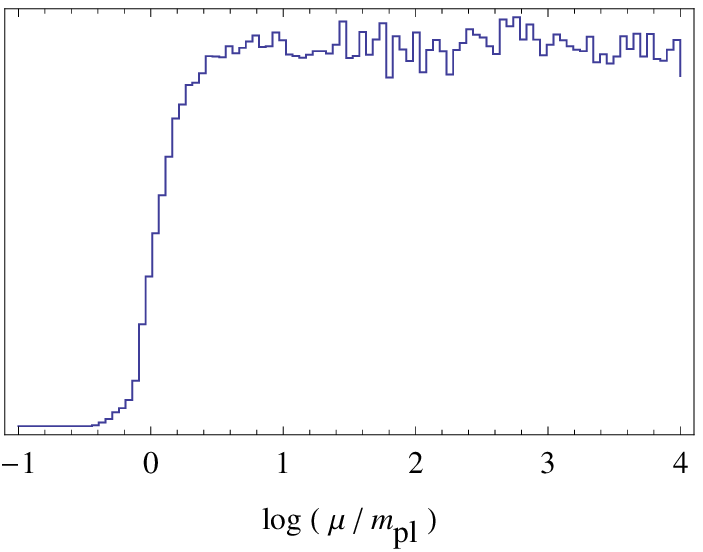}
\includegraphics[width=6cm]{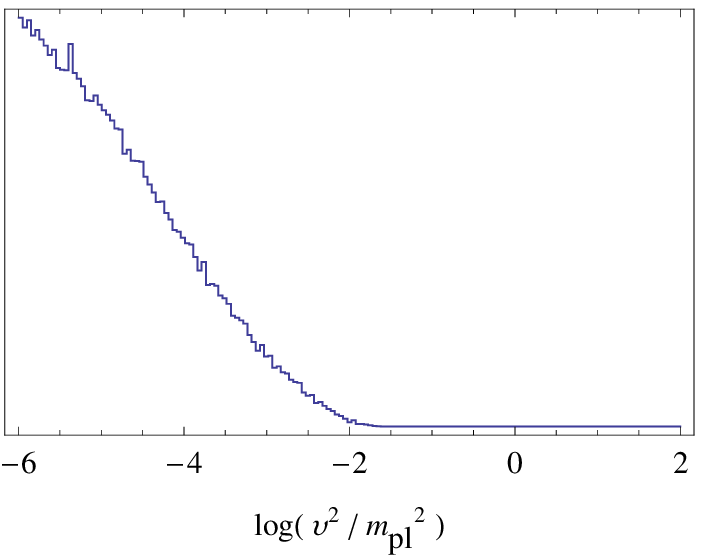}
\caption{Marginalised probability distribution for the potential
  parameters of the hybrid model. Notice that some of the bound are
  set by the prior choices.}
\label{fig:1Dpot}
\end{figure}

The marginalised probability distributions for the potential
parameters are represented in Fig.~\ref{fig:1Dpot}. These
posteriors seem to indicate that the three parameters are bounded but
one should pay attention to the influence of our prior choices over
the posterior~\cite{Trotta:2008bp}. In fact, the posteriors for $M$
and $\nu$ are found to depend on our prior choice: changing the upper
or lower limit on the $\nu$-prior (or $M$-prior) affects the values at
which the $M$- and $\nu$-posteriors fall off. Such a situation is
typical of the existence of correlations between these two
parameters. We have therefore computed the
two-dimensional posterior distribution in the plane $(\nu^2,M)$ and 
found out that this
probability distribution clearly exhibits a correlation between these
two parameters: the lower bound on $M$ depends on the
minimal allowed value of $\nu$ in the prior. Such a correlation comes
from the fact that, to realize enough inflation for a given IC, the 
critical/instability point $\phi_\uc$ should not be larger than the 
anamorphosis image of the IC (the point in the valley where slow-roll 
starts). Restricting initial fields to sub-Planckian values leads to 
an uper bound on the largest image, and thus an upper bound on the 
instability point. From Eq.~(\ref{eq:critical}), the relevant 
quantity that is constrained is the combination $\sqrt{2}
\nu^2 / M =\phi_\uc$.

\begin{figure}
\includegraphics[width=6cm]{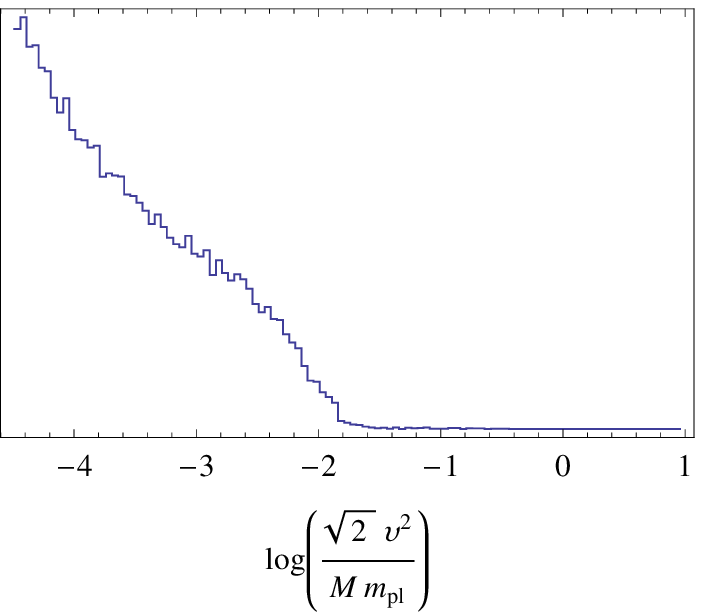}
\caption{Prior independent marginalised posterior probability
  distribution for the parameter $\nu^2/(M \Mpl)$. This parameter
  fixes the position of the instability point and a too large value
  may prevent inflation from occurring in the sub-planckian field
  regime [see Eq.~\ref{eq:critbound})].}
\label{fig:1Dinstpt}
\end{figure}

We have plotted in Fig.~\ref{fig:1Dinstpt} the marginal posterior
distribution associated with the parameter $\log(\sqrt{2}\nu^2/M)$,
and at $95\%$ of confidence level, we find
\begin{equation} 
\label{eq:critbound}
\dfrac{\sqrt{2} \nu^2}{M} < 4\times 10^{-3}.
\end{equation}

The parameter $\mu$ is the other constraint that the MCMC exibits. It
is explained by the possible apparition of slow-roll violations in the
valley, when $\mu$ becomes too small. These slow-roll violations
prevent the generation of an inflationary phase if the trajectory
climbs too high in the valley. At a two-sigma level, one has
\begin{equation}
\label{eq:mubound}
\frac{\mu}{\mpl} > 1.7\,.
\end{equation}
This lower limit is equivalent to the upper limit on $m$ observed in
\cite{Clesse:2008pf}: a large inflaton mass induces a fast roll
evolution and requires super-planckian initial conditions to realize
inflation in a chaotic way. Let us stress that these constraints come
only from requiring enough inflation in the hybrid model whatever the
initial field values, velocities, and other potential parameters. In
this respect, the limits of Eqs.~(\ref{eq:critbound}) and
(\ref{eq:mubound}) can be
considered as ``natural''.

To conclude this section, we have shown that inflation is generic in
the context of the hybrid model and we have derived the marginalised
posterior probability distributions of all the parameters such that 60
e-folds of inflation occur. As discussed in the introduction, the
original hybrid model under scrutiny is however a toy model known to
be disfavored by the current CMB data. In this respect, one may wonder
whether our results are peculiar to this model or can be generalised
to other more realistic two field inflationary models. This point is
addressed in the next section in which we have performed a complete
study of the SUGRA F-term hybrid inflation. In that model, the
dynamics depends on only one potential parameter; also constrained by
cosmic strings formation. The challenge will thus be to confront this
constraint to the natural bounds that can be deducted from MCMC
methods by requiring enough e-folds of inflation.

\section{Probability distributions in F-SUGRA inflation}
\label{section:fsugra}

The minimal supersymmetric versions of hybrid inflation are
known as the F-term and D-term inflationary
models~\cite{Dvali:1994ms,Binetruy:1996xj,Halyo:1996pp}, where the
slope of the valley is generated by radiative corrections. The F-term 
model is compatible with the current CMB data since a red spectrum
of the cosmological perturbations is
generic~\cite{Dvali:1994ms,Senoguz:2003zw,Jeannerot:2005mc}. In
addition, this model is more predictive and testable than its non-SUSY 
version since it contains only one coupling constant and one mass 
scale.

\subsection{The model}

In the following, we are analysing the space of initial conditions and
model parameters leading to enough inflation for the so-called F-term
model based on the superpotential~\cite{Dvali:1994ms}
\begin{equation}\label{superpotFterm}
W_{\rm infl}^{\rm F}=\coupling S(\Phi_+\Phi_- - M^2)\,.
\end{equation}
The inflaton is contained in the superfield $S$. The Higgs pair
$\Phi_+,\Phi_-$ is charged under a gauge group $G$, that is broken at
the end of inflation when the Higgs pair develop a non-vanishing
expectation value (\vev) $M$. The superpotential leads in global SUSY
to a tree level potential
\begin{equation}
V^{\rm SUSY}_{\rm tree}(s,\psi)=\coupling^2\left(M^2-\frac{\psi^2}{4}\right)^2 
+\frac{1}{8}\coupling^2 s^2 \psi^2~,
\end{equation}
where the effective inflaton $s$ and Higgs field $\phi$ can be made
real and canonically normalized [$s \equiv \sqrt{2}\,\Re(S)$,
$\phi=2\Re(\phi_+)=2 \Re(\phi_-)$]. The local minima
of the potential at large $S$ provide a flat direction for the
inflaton $s$: $V_0=\coupling^2M^4$.

This tree level flat direction is lifted by two effects. Firstly,
radiative corrections are induced by the SUSY breaking that supports
inflation. In addition, if the field values are close to the reduced
Planck mass $\Mpl$, one should expect supergravity corrections
$S/\Mpl$ to the tree level potential. The radiative corrections along
the inflationary valley can be derived using the Coleman-Weinberg
formula~\cite{Coleman:1973jx}. They reduce to~\cite{Dvali:1994ms}
\begin{equation}\label{VFterm1dim}
\begin{split}
V^{\rm cw}_{\rm 1-loop}(s)=\frac{\coupling^4M^4\mathcal{N}}{32\pi^2}&\left[2\ln\frac{s^2
\coupling^2}{\Lambda^2}+(z+1)^2\ln(1+z^{-1})\right.\\
&+(z-1)^2\ln(1-z^{-1})\Big],
\end{split}
\end{equation}
where $z=s^2/M^2$, $\mathcal{N}$ stands for the dimensionality of the
representations to which $\Phi_\pm$ belong and $\Lambda$
represents a non-physical energy scale of renormalization. Realistic
values of $\mathcal{N}$ can be derived from the embedding of the model in
realistic SUSY Grand Unified Theories (GUT) as shown in
Ref.~\cite{Jeannerot:2003qv}. For example, in the case of an embedding
of the model in SUSY SO(10), $\Phi_\pm$ belong to the representation
$\mathbf{16},\mathbf{\overline{16}}$ or
$\mathbf{126},\mathbf{\overline{126}}$. However, as pointed out in
Ref.~\cite{Jeannerot:2006jj}, it is possible that only some components
of $\Phi_\pm$ take a mass correction of order $M$ so that
effectively\footnote{This depends on the mass spectrum of the assumed
GUT model.} $\mathcal{N}=2,3$. For the sake of generality, we will 
assume that $\mathcal{N}$ can take values in the range $[2,126]$. This 
model is also known to generically produce cosmic strings at the end of 
inflation \cite{Jeannerot:2003qv} and this imposes an upper limit on the 
inflationary mass scale~\cite{Rocher:2004et,Jeannerot:2005mc,Jeannerot:2006jj}
\begin{equation}
\label{contrainteparamFterm}
M\lesssim 2\times 15\,\mathrm{GeV}, \quad \coupling \lesssim 7\times 10^{-7}\,
\frac{126}{\mathcal{N}}\,.
\end{equation}

Secondly, SUGRA corrections also contribute to lifting the tree-level 
flat direction and will be taken into account since the field values we
are probing are not always negligible compared to the Planck
mass. It has been noticed in Ref.~\cite{Copeland:1994vg} that the
F-term hybrid inflation model doesn't suffer from the $\eta$-problem
only when the K\"ahler potential is (close to) minimal\footnote{We
  will restrict ourselves to minimal SUGRA corrections, neglecting
  SUSY breaking soft terms and the non-renormalizable corrections to
  the superpotential (see \cite{Senoguz:2003zw,Jeannerot:2005mc,Jeannerot:2006jj} 
  for an analysis of their effects).}
\begin{equation}
K\simeq |S|^2+|\Phi_+|^2+|\Phi_-|^2~,
\end{equation}
which is what we assume in the following. In terms of the canonically
normalized effective inflaton $s$ and waterfall fields $\psi$, the
SUGRA corrected potential now reads
\begin{equation}\label{VFSUGRA}
\begin{split}
V_{\rm tree}^{\rm{sugra}}(s,\psi) & = \coupling ^2 \exp \left(\frac{s^2 + \psi^2}{2 \Mpl^2}\right) \\
& \times \Bigg\{ \left( \frac {\psi^2}{4} - M^2 \right)^2  
\left( 1 - \frac{s^2}{2 \Mpl^2} + \frac{s^4}{4 \Mpl^4} \right) \\
&+ \frac{s^2 \psi^2}{4 } \left[ 1+ \frac 1 {\Mpl^2} \left(\frac 1 4 \psi^2 - M^2\right)\right]^2\Bigg\}~.
\end{split}
\end{equation}

The dynamics along the inflationary valley is driven by the radiative
corrections and by the SUGRA corrections. The radiative corrections
play a major role in the last e-folds of inflation (thereby generating
the observable spectral index), whereas most of the dynamics takes
actually place for field values dominated by the SUGRA corrections. We
have calculated the amplitudes for both corrections and found that
only at the end of the inflationary potential (for $s\in [M,8M]$ if
$\mathcal{N}=3$ and $s\in [M,3.5M]$ if $\mathcal{N}=126$), the
radiative corrections may dominate over the SUGRA corrections. In the
present work, the regions of the parameter space leading to inflation
do not depend on the very last part of the field evolution: as soon as
$60$ e-folds are obtained, the initial conditions are considered
successful and this generically occurs in the valley at larger field
values. Outside the inflationary valley, we therefore expect the tree
level dynamics to dominate over the radiative corrections, especially
for small coupling $\coupling$. There also, in addition to the tree
level, at large fields, SUGRA corrections are expected to be
important.

Resulting of these considerations, we have neglected radiative
corrections and used for our study below the potential of 
Eq.~(\ref{VFSUGRA}).

\subsection{Fractal initial field values}

\begin{figure}[] 
\includegraphics[width=8.7cm]{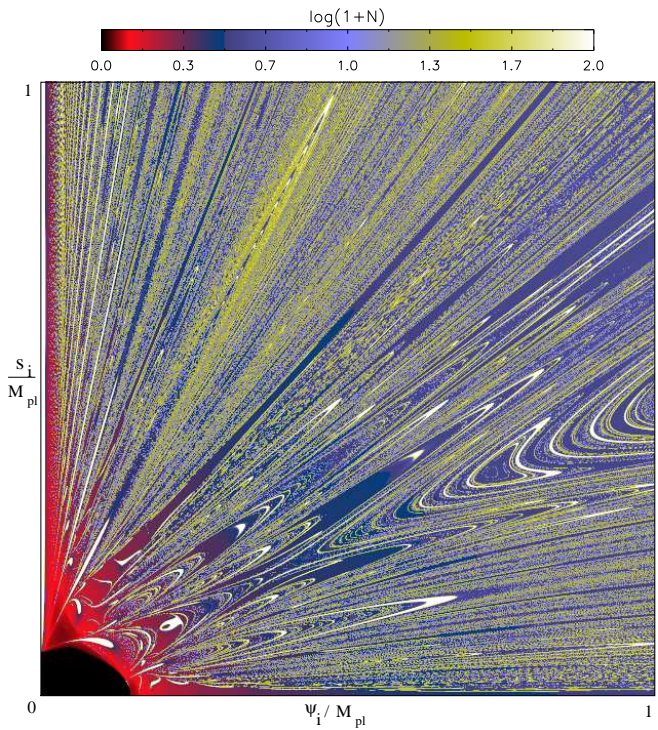}
\caption{Mean number of e-folds (truncated at $100$) obtained from
  $512^2$ initial field values $(\psi_\ui/\Mpl,s_\ui/\Mpl)$ for the
  SUGRA F-term model. The initial field velocities are assumed to
  vanish and the potential parameter is fixed at $M=10^{-2}\mpl$. As
  for the original hybrid model, the mean is computed from $2048^2$
  trajectories (see Fig.~\ref{fig:grid}). The set of initial field
  values producing enough inflation is again of dimension two with a
  fractal boundary.}
\label{fig:gridSUGRA}
\end{figure}

The analysis of the SUGRA F-term model of inflation has been conducted
along the lines described in Sec.~\ref{sec:fractalic} and
Sec.~\ref{section:mcmc}. We have first verified that, at fixed
potential parameter $M$ and vanishing initial velocities, the set of
initial field values $\calS$ defined by Eq.~(\ref{eq:Sdef}) is
two-dimensional with a fractal boundary. In Fig.~\ref{fig:gridSUGRA},
we have represented the set $\calS$ of successful initial field
values for the mass scale $M=10^{-2}\mpl$. Notice that the coupling
constant $\coupling$ being an overall factor, it doesn't impact the
dynamics of the fields. Our study is therefore valid for any value of
$\coupling$ and of the dimensionality of the Higgs field $\mathcal{N}$, 
since the relationship $M(\coupling)$ depends only on $\mathcal{N}$.

As for the original hybrid model, the highest Lyapunov exponent for
the successful initial field values is negative and the set $\calS$ is
non-chaotic. Outside of $\calS$, trajectories have positive Lyapunov
exponents and exhibit chaos.

\begin{table}
\begin{center}
\begin{tabular}{|l|l|}
\hline Values of $M$ & Area of $\calS$ (\%)
\\\hline\hline
$M=10^{-1}~\mpl$  &   0 (exact)\\
$M=10^{-2}~\mpl$ & $12.9 \pm 0.1$\\ 
$M=10^{-3} \mpl $ & $12.0 \pm 0.3$ \\ 
$M=10^{-4} \mpl $ & $10.3 \pm 0.5$ \\ 
\hline
\end{tabular}
\caption{Percentage of successful initial field values, at vanishing
  initial velocities, for various values of the potential parameter
  $M$. The error bars come from the finite numerical precision, 
  which decreases with $M$.}
\label{tab:successFterm}
\end{center}
\end{table}

For vanishing initial velocities, we have reported in
Table~\ref{tab:successFterm} the area occupied by the set $\calS$ in
the plane $(s_\ui,\psi_\ui)$ for various section along the potential
parameter $M$. Like for the original hybrid model, we recover a
significant proportion of successful initial field values outside the
valley. This result holds even for $M\ll 1$ though at small M, the
potential becomes very flat and the number of oscillations of the
system before being trapped in the inflationary valley can exceed
$10^3$. Simulations become therefore more time-consuming and
error-bars in Tab.~\ref{tab:successFterm} increase. Reducing $M$ also
reduces the typical size of structures in the plane
$(s_\ui,\psi_\ui)$, which evolves from Fig.~\ref{fig:gridSUGRA} to a
more intricated space of thinner successfull IC. As suggested by the
Tab.~\ref{tab:successFterm}, we will see below that this doesn't
affect the probability of getting inflation by starting the field
evolution outside the valley.

\begin{figure}[]
\includegraphics[width=8cm]{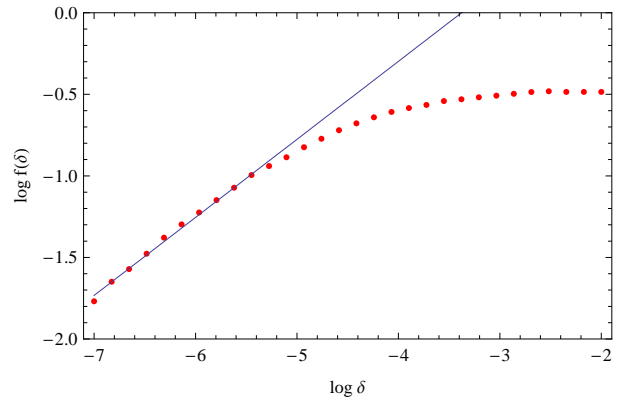}
\caption{Fraction of initial field values in a $\delta$-sized box
    intercepting the set $\calS$ as a function of $\delta$ for the
    SUGRA F-term model. The potential parameter has been fixed to
    $M=10^{-2} \mpl$. The box-counting dimension of boundary of 
    $\calS$ is
    given by the power law behaviour for small $\delta$ and found to
    be $D_\uB\simeq 1.5$.}
\label{fig:dim_fractale_SUGRA}
\end{figure}

Concerning the fractal properties of $\calS$, we have applied the same
method as in Sec.~\ref{sec:boxcountingdim} to compute the box-counting
dimensions of $\calS$ and its boundary. As expected, we recover that
$\calS$ is of box-counting dimension two whereas the function
$f(\delta)$ for its boundary is represented in
Fig.~\ref{fig:dim_fractale_SUGRA}. We obtain that, as in the non-SUSY
case, the boundaries are fractal with dimension
\begin{equation}
D_\uB \simeq 1.5~.
\end{equation}
These results allow us to use the usual Lebesgue measure to define
the probability distribution over the whole parameter space.

\subsection{MCMC on the initial field values, velocities and the potential 
parameter}

\begin{figure}
\includegraphics[width=9cm]{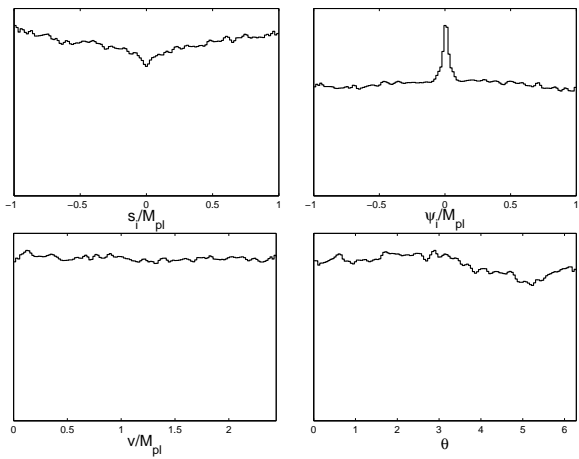}
\caption{Marginalised posterior probability distributions for the
  initial field values (upper panels) and the initial velocities,
  modulus $v$ and angle $\theta$. The F-SUGRA inflationary valley has
  a slightly higher probability density around $\psi=0$ but is
  extremely localised: as a result, inflation is more probable by
  starting out of the valley.}
\label{fig:fsugra1Dfields}
\end{figure}

As already mentioned, there is only one potential parameter $M$ in
F-term SUGRA model that may influence the two field dynamics. The goal
of this section is to evaluate the probability distributions of the
initial field values, velocities and of $M$ such that inflation lasts
more than $60$ e-folds. As for the original hybrid model, we have
performed an MCMC analysis on the five-dimensional parameter space
defined by $s_\ui$, $\psi_\ui$, $v$, $\theta$ and $M$ where
\begin{equation}
\left. \dfrac{\ud s}{\ud N}\right|_\ui = v \cos \theta, \quad
\left. \dfrac{\ud \psi}{\ud N}\right|_\ui = v \sin \theta.
\end{equation}

We have chosen the same sub-planckian priors for the initial field
values and initial velocities than in Sec.~\ref{section:mcmc}. Since
the order of magnitude of $M$ is not known, we have chosen a flat
prior on 
\begin{equation}
  - 2 < \log \dfrac{M}{\Mpl} < 0.
\end{equation}
The lower limit on $M$ is motivated by computational rather than
physical considerations. The resulting marginalised posterior
probability distributions for each of the parameters are represented
in Fig.~\ref{fig:fsugra1Dfields}. The chains contain $400000$ samples
producing an estimated error on the posteriors around a few percent
(from the variance of the mean values between different chains).

\begin{figure}
\includegraphics[width=9cm]{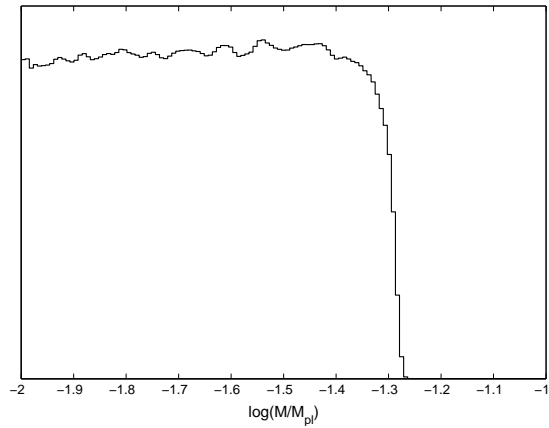}
\caption{Marginalised posterior probability distribution of the mass
  scale $M$ of F-SUGRA inflation.}
\label{fig:fsugraM}
\end{figure}

The posteriors for the field velocities are flat showing that all
values are equiprobable to produce inflation. The initial field values
are also flat, up to a sharp peak of higher probability density around
$\psi=0$ corresponding to the inflationary valley. As for the hybrid
model of Sec.~\ref{sec:fractalic}, after integration of these curves
over the field values, inflation is clearly more probable by starting
out of the valley. Finally, only the posterior probability
distribution of $\log M$ is strongly suppressed at large values. We
find, at $95\%$ of confidence level (see Fig.~\ref{fig:fsugraM}).
\begin{equation}\label{eq:M2sig}
\log(M) < -1.33\,.
\end{equation}
As for the original hybrid model, this limit comes from the condition
of existence of a sub-planckian inflationary valley which is related
to the position of the instability point. Indeed, from
Eq.~(\ref{VFSUGRA}), one finds
\begin{equation}
  \left. \dfrac{\ud V^{\mathrm{SUGRA}}_{\mathrm{tree}}}{\ud
      \psi}\right|_{\psi=0} = 0 \Rightarrow s=s_\uc=\pm \dfrac{M}{\Mpl}\sqrt{1-\sqrt{1-4\dfrac{M^4}{\Mpl^4}}}\,,
\end{equation}
where we have kept only the sub-planckian solutions. This expression
shows that there is an inflationary valley at $\psi=0$ only for $M/\Mpl
< 1/\sqrt{2}$, and for field values such that $s>s_\uc$. As a result of
the two-field dynamics, we find that a valley supporting at least $60$
e-folds of inflation require the more stringent bound of
Eq.~(\ref{eq:M2sig}). Let us finally notice that the most probable
values we obtain on $M$ to get inflation in Eq.~(\ref{eq:M2sig}) are
compatible with the existing upper bound coming from cosmic strings
constraint: $M \lesssim 10^{-3} \mpl$ (see
Ref.~\cite{Rocher:2004et,Jeannerot:2005mc}).

\section{Conclusion}

\label{section:conclusion}

In this paper, by numerically solving the two-field dynamics of the
original hybrid model and its SUGRA F-term version, we have shown that
$60$ e-folds of inflation is a generic outcome. Contrary to what is
usually assumed, one does not need to fine-tune the initial field
values around $\psi=0$ to get inflation. In fact, the inflationary
valley, indeed of small extension in field space, is one of the three
dynamical attractors of the differential system given by the Einstein
and Klein--Gordon equations in a FLRW universe (the others being the
minima of the potential). As a result, any trajectory will end in one
of these three attractors and the set $\calS$ of successful initial
conditions therefore belongs to the basin of attraction of the
inflationary valley. We have shown that such a set is connected and of
dimension two while exhibiting a fractal boundary of dimension greater
than one. Moreover, it occupies a significant fraction of the
sub-planckian field regime. In order to quantify what are the natural
field and parameter values to get inflation for both of these models,
we have introduced a probability measure and performed a MCMC
exploration of the full parameter space. It appears that the
inflationary outcome is independent of the initial field velocities,
is more probable when starting out of the inflationary valley, and
favours some ``natural'' ranges for the potential parameter values
that cover many order of magnitudes. The only constraints being that
the inflationary valley should at least exist.

Let us notice that the posterior probability distributions we have
derived are not sensitive on the fractal property of the boundary of
$\calS$. This is expected since, even fractal, the boundary remains of
null measure compared to $\calS$. However, its existence may have
implications in the context of chaotic eternal
inflation~\cite{Linde:1986fd, Guth:2000ka}. Indeed, the boundary
itself leads to inflation and spawn the whole field space such that
its mere existence implies that inflationary bubbles starting from
almost all sub-planckian field values should be produced. Here, we
have been focused to the classical evolution only and our prior
probability distributions have been motivated by theoretical prejudice
(flat sub-planckian prior). In the context of chaotic eternal
inflation, our results are however still applicable provided one uses
the adequate prior probabilities which are the outcome of the
super-Hubble chaotic structure of the
universe~\cite{Linde:2008xf}. Provided the eternal scenario does not
correlate with the classical dynamics, one should simply factorise the
new priors with the posteriors presented here to obtain the relevant
posterior probability distributions in this context.

\begin{acknowledgments}
  It is a pleasure to thank T. Carletti, S. Colombi,
  G. Esposito-Farese, A. Fuzfa, A. Lemaitre, J.~Martin, P. Peter, and
  M. Tytgat for discussions and comments. S.C. is supported by the
  Belgian Fund for research (F.R.I.A.). J.R. is funded in part by IISN
  and Belgian Science Policy IAP VI/11.
\end{acknowledgments}

\bibliography{biblio}

\end{document}